\documentclass[journal=nalefd,manuscript=letter,layout=twocolumn,maxauthors=10]{achemso}
\usepackage{graphicx}
\usepackage{epsfig}
\usepackage{dcolumn}


\setkeys{acs}{articletitle = true,etalmode=truncate,maxauthors=10} 
\usepackage{amsmath}

\usepackage{bm}
\usepackage{rotating}
\usepackage{multirow}
\usepackage[table]{xcolor}
\usepackage[english]{babel}

\def\ve{{\varepsilon}}
\def\w{\omega}
\def\bk{{\bf k}}

\def\bq{{\bf q}}

\def\>{\rangle}
\def\<{\langle}
\def\D{\partial}

\title{Nonequilibrium Lattice Dynamics in Monolayer MoS$_2$}

\author{Fabio Caruso}
\email{caruso@physik.uni-kiel.de}
\affiliation{Institut f\"ur Theoretische Physik und Astrophysik, Christian-Albrechts-Universit\"at zu Kiel, Kiel, Germany} 
\begin{document}

\maketitle

\begin{abstract}
The coupled nonequilibrium dynamics of electrons 
and phonons in monolayer MoS$_2$ is investigated
by combining first-principles calculations of the  
electron-phonon and phonon-phonon interaction with 
the time-dependent Boltzmann equation. 
Strict phase-space constraints in the electron-phonon 
scattering are found to influence profoundly the decay path of excited 
electrons and holes, restricting the emission of phonons to 
crystal momenta close to few high-symmetry points in the Brillouin zone.
As a result of momentum selectivity in the phonon emission, 
the nonequilibrium lattice dynamics is characterized by the emergence of  
a highly-anisotropic population of phonons in reciprocal space,  
which persists for up to 10~ps until thermal 
equilibrium is restored by phonon-phonon scattering. 
Achieving control of the nonequilibrium dynamics of the lattice
may provide unexplored opportunities to selectively 
enhance the phonon population of two-dimensional crystals and, thereby, 
transiently tailor electron-phonon interactions 
over sub-picosecond time scales.
\end{abstract}

Transition-metal dichalcogenides (TMDs)
exhibit strong light-matter coupling,\cite{PhysRevLett.105.136805}
a rich photo-physics,\cite{mueller_exciton_2018}
and a unique interplay of spin, \cite{zhou_spin-orbit_2019,Molina2013}
valley, \cite{PhysRevLett.123.236802,PhysRevLett.124.217403}
lattice,\cite{kang_holstein_2018} and electronic \cite{PhysRevLett.111.216805} degrees of freedom. 
These characteristics, alongside with numerous 
possibilities to tailor screening, \cite{briggs_atomically_2020} charge-carrier density, \cite{kang_holstein_2018} 
and dimensionality, \cite{kang_high-mobility_2015,Cuadra2018} make them promising candidates for the 
exploration of new pathways to achieve properties on demand 
in quantum matter.\cite{basov_towards_2017}
A broad spectrum of emergent phenomena can indeed be realized 
and controlled in these compounds via coupling to femtosecond 
light pulses, including  
phase transitions,\cite{ultrafast/melting/Rossnagel/2010,hellmann_time-domain_2012} 
transient band-structure \cite{Rossnagel2014,hein_mode-resolved_2020} and 
band-gap renormalization,\cite{chernikov_population_2015,doi:10.1021/acsnano.5b06488,doi:10.1021/acs.jpclett.0c00169}
and tunable valley polarization.\cite{berghauser_inverted_2018,PhysRevLett.123.236802}
While these phenomena are a manifestation of complex 
and diverse interaction mechanisms between electrons 
and lattice, a detailed understanding of the 
lattice dynamics  and electron-phonon coupling
in systems out of equilibrium remains elusive, 
and it keeps providing a strong stimulus for 
theoretical and experimental condensed-matter research.\cite{ultrafast/melting/Rossnagel/2010,PhysRevLett.123.236802,doi:10.1021/acs.nanolett.6b04419,
hein_mode-resolved_2020,PhysRevResearch.1.022007,doi:10.1021/acs.nanolett.7b00175} 

Time- and angle-resolved photoemission spectroscopy (tr-ARPES)
has led the way in the experimental investigation of the  
out-of-equilibrium carrier dynamic in TMDs, \cite{doi:10.1021/acs.nanolett.5b01967}
providing direct insight on the influence of electron-phonon interactions on the carrier dynamics.\cite{Chen21962}
Information regarding the lattice and its nonequilibrium 
dynamics, however, can only be inferred indirectly
from its effects on the electrons, as, e.g., 
transient changes of the Fermi level, electron 
binding energies, 
or quasi-particle linewidths.\cite{Schmitt1649,Rossnagel2014,hein_mode-resolved_2020}
Femtosecond electron diffuse scattering\cite{Sciaini_2011} (FEDS)
probes the progressive change of phonon population
as the dynamics of the coupled electron-phonon system evolves,\cite{Siwick2019}
and it thus offer a more direct and detailed picture of the out-of-equilibrium
dynamics of the lattice.\cite{Siwick1382}
Recent FEDS measurements provide strong evidence that 
the thermalization of electronic and vibrational degrees of freedom in TMDs 
is characterized by striking anisotropies in the phonon 
population in reciprocal space.\cite{PhysRevLett.119.036803,otto2020mechanisms}
These manifest themselves through the emergence of hot-spots 
in the FEDS intensity, which indicates local enhancement of the 
phonon population at selected high-symmetry points and 
directions in the Brillouin zone.\cite{Stern2018,Siwick2019,seiler2020accessing,
doi:10.1021/acs.nanolett.9b01179,seiler2020accessing,otto2020mechanisms}

A quantitative understanding of these phenomena is challenging owing 
to the combined influence of valley degrees of freedom, lattice 
anharmonicities, and anisotropies of the electron-phonon interaction 
on the lattice dynamics. 
Despite providing valuable insight into the carrier dynamics of two-dimensional 
and layered materials,\cite{bib:johannsen13,Bauer/PRB15,Waldecker2016} 
the concept of {\it hot phonon} 
-- whereby the hot-carrier relaxation is assumed to be dominated 
by a few strongly-coupled phonon modes 
\cite{Allen1987,bib:lin08,PhysRevLett.122.016806,novko2020,Caruso2020} -- 
is unsuitable for the description of these phenomena, since it 
inherently lacks a momentum-resolved description of the lattice dynamics. 
In the domain of first-principles calculations, conversely, 
the Boltzmann equation formalism \cite{Ziman:100360,Darancet,doi:10.1021/acs.nanolett.7b02212} 
is emerging as a promising technique for investigating the 
influence of electron-phonon interactions on the 
nonequilibrium dynamics of electrons and phonons.

In this manuscript, the coupled nonequilibrium dynamics of electrons 
and phonons in monolayer MoS$_2$ is investigated from first principles  
based on the time-dependent Boltzmann equation. 
Many-body effects due to the electron-phonon 
and phonon-phonon interactions are explicitly 
accounted for within a new time-propagation algorithm, which enables the
investigation of the nonequilibrium lattice dynamics with unprecedented 
resolution in reciprocal space. 
Inter-valley scattering processes in the relaxation path of 
excited carriers are found to influence thoroughly the 
lattice dynamics of monolayer MoS$_2$, confining the 
phonon emission to the $\Gamma$ and K high-symmetry points in the 
Brillouin zone. This mechanism is responsible for 
establishment of a non-thermal regime
in the vibration of the crystalline lattice, which is characterized by a pronounced 
momentum anisotropy in the phonon population and is found to 
persist for several picoseconds.
Overall, these findings suggest a new route to transiently 
control the phonon population, the electron-phonon coupling, 
and their nonequilibrium dynamics over timescales of several picoseconds. 

\begin{figure*}[t]
\begin{center}
   \includegraphics[width=0.68\textwidth]{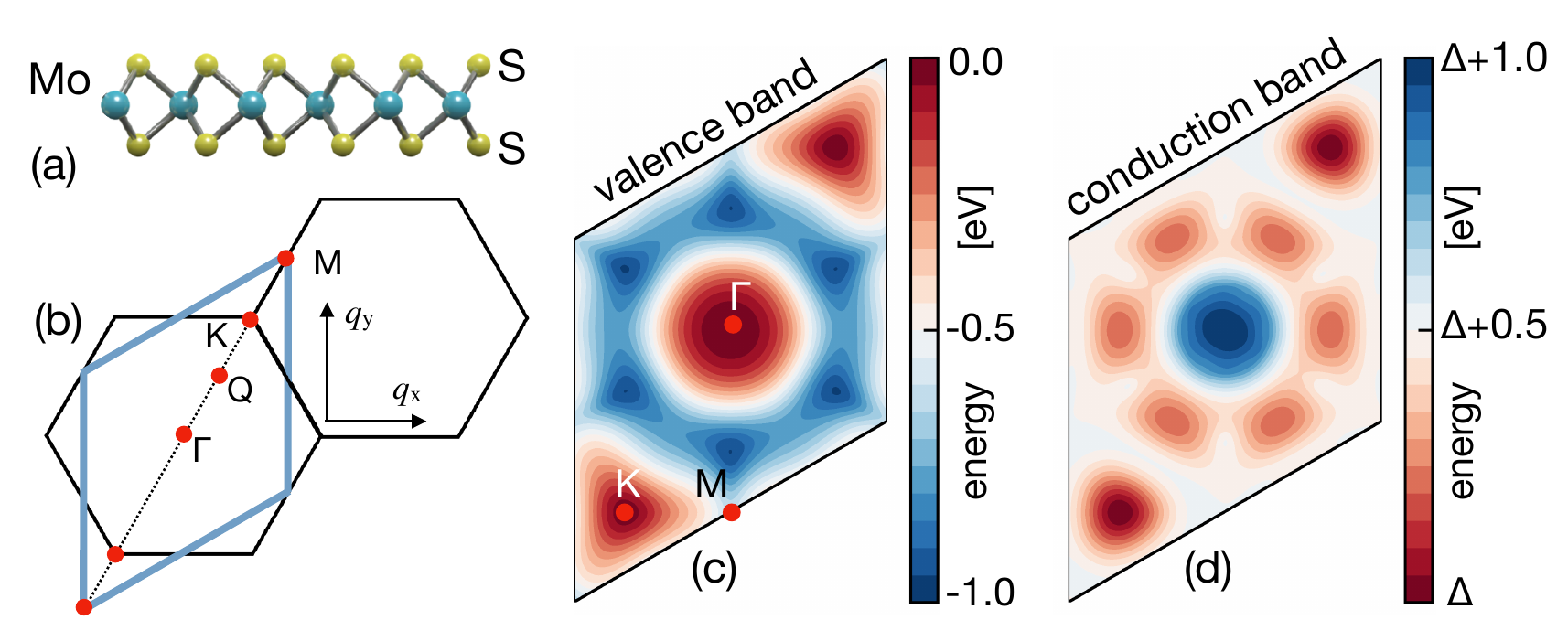}
\caption{\label{fig:bands} 
(a) Side view of the monolayer MoS$_2$ crystal structure. 
(b) Brillouin zone and high-symmetry points. 
The $\Gamma$-centered rhomboidal BZ is marked in blue. 
Energy dispersion of the valence (c) and conduction bands (d) in the BZ, as obtained
from density-functional theory. 
Energies are relative to the valence band maximum, and $\Delta=1.7$~eV is 
the energy of the Kohn-Sham band gap.}
\end{center}
\end{figure*}

At thermal equilibrium, 
the occupation of electronic and vibrational energy levels is 
given by the Fermi-Dirac and Bose-Einstein distribution functions, respectively: 
\begin{align}
f_{n\bk}^{\rm FD}   (\mu,T) &= [e^{(\ve_{n\bk}- \mu)/k_{\rm B}T}+1 ]^{-1} \label{eq:FD} \\
n_{\bq\nu}^{\rm BE} (T)     &= [e^{\hbar\w_{\bq\nu} /k_{\rm B}T}-1 ]^{-1} \label{eq:BE}. 
\end{align}
Here, $\ve_{n\bk}$ is the single-particle energy of a Bloch state, 
$\mu$ the chemical potential,  $\hbar\w_{\bq\nu}$ the phonon energy, 
$k_{\rm B}$ the Boltzmann constant, and $T$ the temperature. 
Out of equilibrium, 
either $f_{n\bk}$ or $n_{\bq\nu}$ differs
from Eqs.~\eqref{eq:FD} and \eqref{eq:BE}, 
and the dynamics of the coupled electron-phonon system
is driven by scattering processes between electrons and phonons. 
The time dependence of the distribution functions may be described within the 
Boltzmann equation formalism:\cite{haug2007quantum,Darancet,doi:10.1021/acs.nanolett.7b02212}
\begin{align}
\label{eq:bte_f}  {\D_t f_{n\bk}(t)}     &= \Gamma^{\rm ep}_{n{\bk}}[f_{n\bk}(t), n_{\bq\nu}(t)] \\ 
\label{eq:bte_n}  {\D_t n_{\bq \nu}(t)}  &= \Gamma^{\rm pe}_{{\bq\nu}}[f_{n\bk}(t), n_{\bq\nu}(t)]  + \Gamma^{\rm pp}_{{\bq\nu}}[n_{\bq\nu}(t)]
\end{align}
where ${\D_t} = \D / \D t$.
The interplay of electronic and nuclear degrees of freedom is 
accounted for by the collision integrals for the electron-phonon ($\Gamma^{\rm ep}_{n{\bk}}$), 
and phonon-electron ($\Gamma^{\rm pe}_{{\bq\nu}}$) interaction respectively, whereas $\Gamma^{\rm pp}_{{\bq\nu}}$ 
accounts for phonon-phonon scattering processes induced by lattice anharmonicities. 
Explicit expressions for the collision integrals are 
reported in the Supporting Information.  
Momentum-resolution in the time-evolution of the electron distribution function is
treated approximately by the introduction of a density-of-state approximation.\cite{doi:10.1063/1.4961874,Sjakste_2018}
Radiative carrier recombination in MoS$_2$ occurs on time 
scales of the order of 1~ns \cite{Amani1065,PhysRevB.93.201111} 
or longer,\cite{Bataller2019} and it is hereby assumed to be 
inconsequential for the lattice dynamics. 
Electron-hole interactions are also neglected. 
Recent studies indicated that non-adiabatic effects \cite{PhysRevLett.119.017001}
may lead to a renormalization of the optical-phonon
energy by up to 2~meV in doped MoS$_2$ \cite{PhysRevX.9.031019,novko_broken_2020,PhysRevB.101.054304}.
These effects have been neglected here.
To investigate the coupled nonequilibrium 
electron-phonon dynamics of monolayer MoS$_2$, 
Eqs.~\eqref{eq:bte_f} and \eqref{eq:bte_n} have been solved 
by time-stepping the time derivative for a total duration of 
40~ps using a time step of 1~fs. 
Further details are found in the Supporting Information.

\begin{figure*}[t]
   \includegraphics[width=0.98\textwidth]{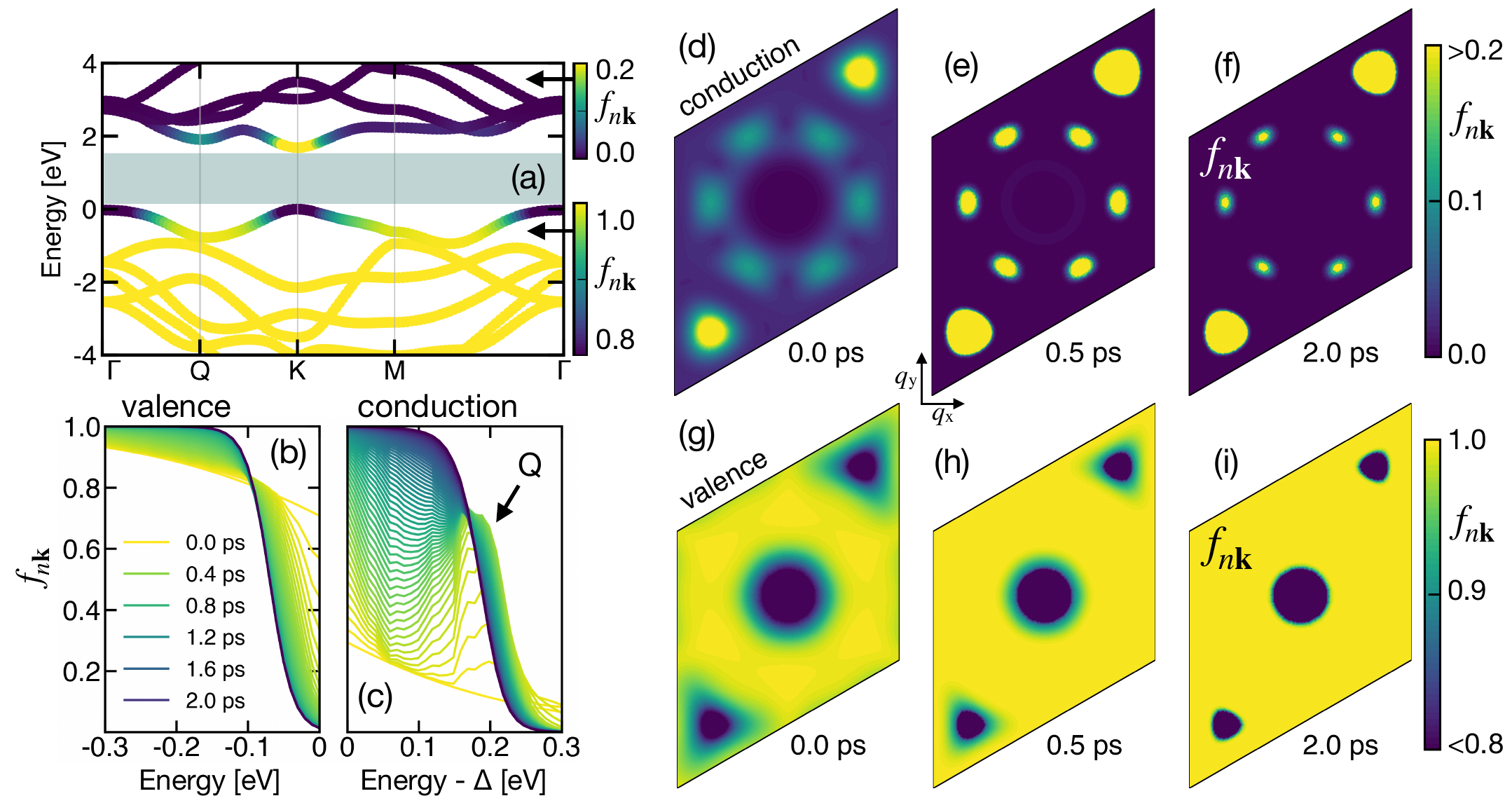}
\caption{\label{fig:fig_f}
Nonequilibrium dynamics of electrons and holes in monolayer  MoS$_2$.
(a) DFT band structure and 
Fermi-Dirac occupations (superimposed as a color coding) for an initial  
excited electronic state $f_{n{\bf k}}(t=0)$. 
The band gap is shaded and different color scales are used for conduction and valence states, respectively. 
Time and energy dependence of the electron distribution function 
$f_{n{\bf k}}$ in the valence (b) and conduction (c) bands. 
The conduction-band energy is relative to the energy 
of the Kohn-Sham band gap ($\Delta = 1.7$~eV). 
Time- and momentum-resolved electronic distribution function $f_{n\bk}$ 
for crystal momenta in the full BZ for the conduction (d-f) and valence (g-i) bands 
at times $t=0$, 0.5, and 2~ps. 
 }
\end{figure*}

A side view of the monolayer MoS$_2$ crystal structure is shown 
in Fig.~\ref{fig:bands}~(a), whereas its 
two-dimensional (2D) hexagonal Brillouin zone (BZ) is reported in 
Fig.~\ref{fig:bands}~(b). In the following, 
the $\Gamma$-centered rhomboidal primitive cell in 
reciprocal space is considered (referred to below simply
as BZ, and marked in blue in Fig.~\ref{fig:bands}~(b)).
For momenta in the BZ, the valence and conduction bands 
are illustrated in Figs.~\ref{fig:bands}~(c) and (d), respectively, whereas
the full band structure on a path is reported in Fig.~\ref{fig:fig_f}~(a).
Close to the gap, the valence band is characterized by 
a non-degenerate valley at $\Gamma$ and a doubly-degenerate valley 
at K, whereas the conduction band has 
a doubly-degenerate valley at K and a higher-energy 
sixfold-degenerate valley at Q. 
The co-existence of several quasi-degenerate valleys 
is shown below to play a key role in the nonequilibrium 
dynamics of the lattice. 

In the following, I focus on a scenario in which the
electron and lattice dynamics is triggered by the
return to equilibrium of an electronic excitation.
The chosen initial state is representative of the 
excitation conditions that can be achieved in 
pump-probe experiments for mono-layer MoS$_2$ \cite{seo_ultrafast_2016}. 
At time $t=0$, it is characterized by a density $n=1 \cdot 10^{14}$~cm$^{-2}$ 
of electrons (holes) promoted to the conduction (valence) bands. 
Electronic occupations are given 
by a Fermi-Dirac distribution function $f_{n\bk}(t=0) =  f^{\rm FD}_{n\bk}(\mu_{\rm e/h},T_{\rm el}^{0})$,  
with an {\it effective} electronic temperature $T_{\rm el}^{0}=2000$~K higher 
than the initial temperature of the lattice $T_{\rm ph}^0=100$~K. 
Other initial conditions are discussed in the Fig.~S1 in the Supporting Information. 
The chemical potential of the electrons in the conduction band, $\mu_{\rm e}$, 
is determined through solution of the integral 
equation $n= {\Omega_{\rm BZ}^{-1}} \sum_m^{\rm cond.} \int {d\bk} 
f^0_{m\bk} ( \mu_{\rm e} , T_{\rm el}^{0})$, where 
$\Omega_{\rm BZ}$ is the area of the 2D Brillouin zone, and the sum extends over the conduction manifold. 
The hole chemical potential $\mu_{\rm h}$ is determined analogously.

The carrier density $n$ and the temperatures $T_{\rm el}^0$ and  $T_{\rm ph}^0$
define the initial conditions for the propagation of 
Eqs.~\eqref{eq:bte_f} and \eqref{eq:bte_n} in time. 
At thermal equilibrium ($T_{\rm el}^0 = T_{\rm ph}^0$),
changes of $f_{n\bk}$ due to the absorption and emission of
phonons are exactly balanced, the right-hand side of Eq.~\eqref{eq:bte_f} vanishes,
and equilibrium is preserved (${\D_t f_{n\bk}(t)}  =0$). The same applies to $n_{\bq \nu}$.
Conversely, if  $T_{\rm el}^0 > T_{\rm ph}^0$ electron-phonon scattering processes 
ensue to reestablish thermal equilibrium. 
The excess energy per unit cell of the initial electronic distribution 
can be estimated via
$\Delta E_{\rm el} = {\Omega_{\rm BZ}^{-1}} \sum_{n}\int \,{d\bk} \, \ve_{n\bk} [f^{\rm FD}_{n\bk }(T_{\rm el}^0) - f^{\rm FD}_{n\bk }(T_{\rm ph}^0) ]$, 
which yields $\Delta E_{\rm el} = 35$~meV per unit cell for the conditions specified above. 

In Fig.~\ref{fig:fig_f}~(a),  
the electron distribution function $f_{n\bk}(t=0)$ 
corresponding to the initial excited state 
is superimposed to the band dispersion of monolayer MoS$_2$. 
Bright regions in the conduction band 
reflect the initial population of excited electrons, whereas
dark regions in the valence band indicate the hole population.
The distribution function $f_{n{\bf k}}^0$ -- further illustrated 
in the full BZ in Fig.~\ref{fig:fig_f}~(d) and (g) for the 
conduction and valence bands, respectively -- indicates that 
excited electrons (holes) primarily occupy states in the vicinity 
of the K and Q (K and $\Gamma$) high-symmetry points.  

\begin{figure*}[t]
   \includegraphics[width=0.98\textwidth]{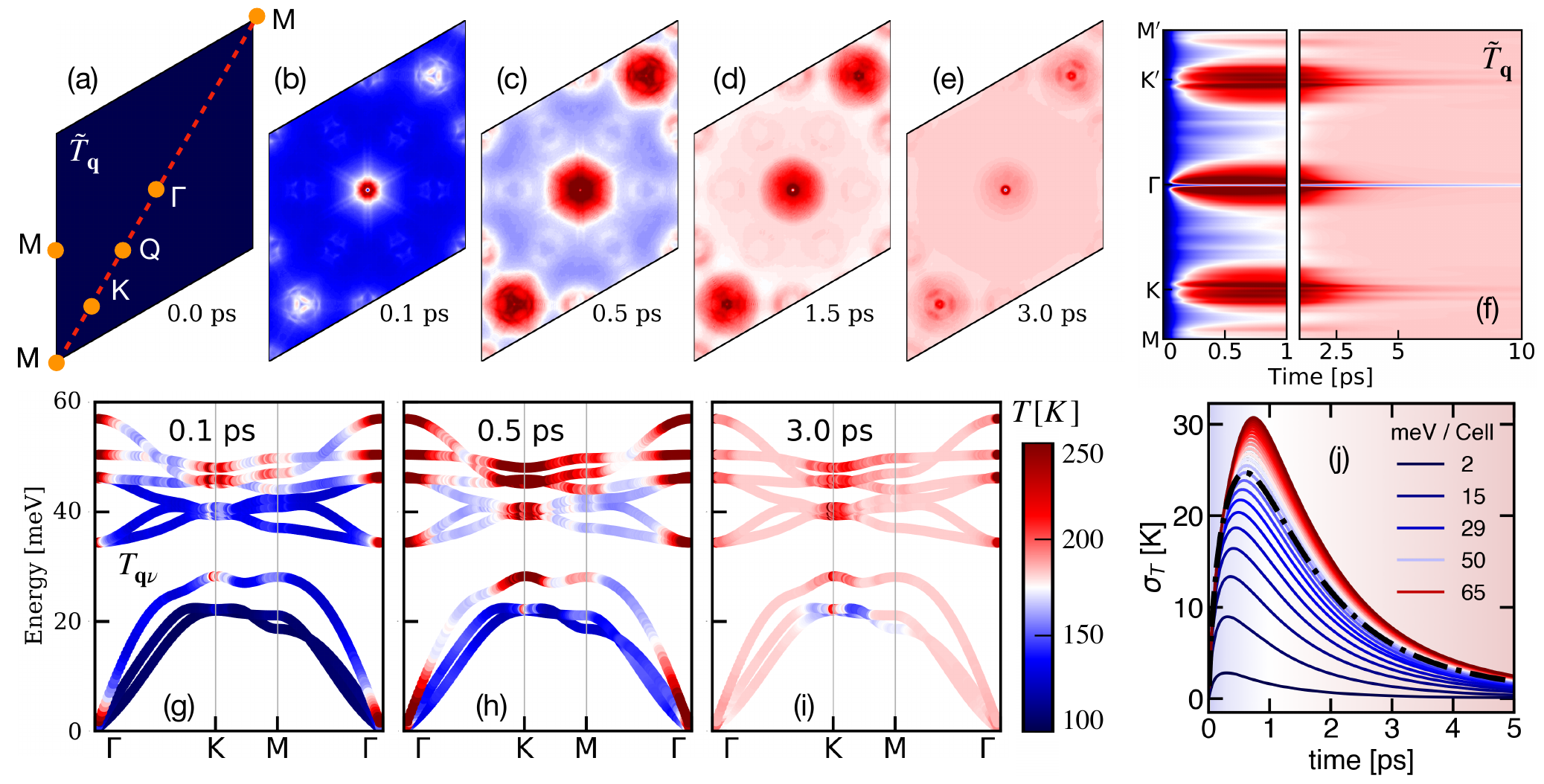}
\caption{\label{fig:fig_n}
Nonequilibrium lattice dynamics of monolayer MoS$_2$.
(a) Momentum-resolved effective phonon temperature $\tilde T_{\bq}$ (defined below Eq.~\eqref{eq:T})
at thermal equilibrium, 
and  (b-e) at several time delays throughout the thermalization process. 
The same colors scale (color bar beside panel (i)) is used 
for panels (a-i).
(f) Time dependence of the effective phonon temperature   $\tilde T_{\bq\nu}$
along the diagonal M-K-$\Gamma$-K-M path 
in the BZ (dashed line in panel (a)). 
(g-i) mode- and momentum-resolved effective phonon temperature 
$T_{\bq\nu}$ (Eq.~\eqref{eq:T}) superimposed to the phonon dispersion 
of MoS$_2$ for $t = 0.1$, 0.5, and 3.0~ps. 
(j) Standard deviation of the effective phonon temperature
(Eq.~\eqref{eq:std}) for initial excitation energies $\Delta E_{\rm el}$ 
ranging between 2 and 65~meV per unit cell. 
}
\end{figure*}

Figures~\ref{fig:fig_f}~(b-c) report 
the electronic occupations $f_{n\bk}$ 
in the valence and conduction bands, 
respectively, throughout the first 2~ps of the dynamics. 
Since radiative recombination is neglected here, 
the total density of excited electrons and holes 
remains constant throughout the dynamics. 
The qualitative agreement between the temperature dependence 
of the Fermi-Dirac function (Fig.~S4 of the Supporting Information) 
and the changes of the electronic occupations suggests 
an intuitive picture of the electron dynamics, whereby thermalization 
is achieved through a progressive lowering of the effective electronic temperature. 
It takes about 800~fs for excited holes in the 
valence band to thermalize with the lattice, 
whereas the electronic relaxation in the 
conduction band is completed within 2~ps.
These time scales are in excellent agreement with 
recent femtosecond electron diffraction 
measurements on monolayer MoS$_2$,\cite{Mannebach/2015}
which estimated to $1.7\pm0.3$~ps the timescale for electronic 
thermalization via electron-phonon scattering, 
whereas relaxation timescales of the order of 1~ps have 
been reported for few-layer samples.\cite{doi:10.1021/nn504760x}
The different time scales for electron and hole relaxation
can be ascribed to the co-existence of three 
quasi-degenerate valleys at $\Gamma$, K, and K$^\prime$ 
in the valence band, which in turn provide for a larger 
phase space for electron-phonon scattering.
As electrons and lattice approach thermal equilibrium, 
$f_{n\bk}$ converges towards a Fermi-Dirac function with 
final temperature $T_{\rm el}^{\rm fin} = 180$~K 
(dark blue in Fig.~\ref{fig:fig_f}~(b-c)). 

Interestingly, while the distribution function $f_{n\bk}$  remains monotonic
in the valence band at each time step, revealing no traces of 
population inversion, 
a transient peak in the electronic occupations of the 
conduction band (arrow in Fig.~\ref{fig:fig_f}~(c)) emerges 
over the first 300~fs at 200~meV above the conduction-band minimum,   
the energy of the six-fold degenerate Q pocket. 
This feature indicates that a bottleneck effect in the 
carrier relaxation may occur at Q, leading to a transient 
accumulation of hot carriers around the Q point, 
and it suggests that, similarly to WS$_2$,\cite{chernikov_population_2015}
a regime of population inversion might be established in monolayer MoS$_2$
under suitable conditions of photo excitation.

A momentum-resolved view of the electron and hole dynamics is given by 
Figs.~\ref{fig:fig_f}~(e-f) and (h-i), where
values of $f_{n\bk}$ in the full BZ are shown for the conduction and valence 
bands at selected time snapshots. 
Throughout the dynamics, excited electrons and holes 
remain localized in momentum space in the vicinity  of 
the K and $\Gamma$ high-symmetry points in the valence band, 
and around K and Q in the conduction band. 
As time evolves, the occupation of electronic states in the BZ
-- initially, more diffused owing to the higher electronic 
temperature -- localizes further in the vicinity  of 
high-symmetry points.
This trend reflects a lowering of the 
electronic temperature as energy is transferred to 
the lattice and carriers scatter back to the Fermi energy. 

Having discussed the dynamics of excited electrons and holes, 
I proceed next to discuss the out-of-equilibrium dynamics of the lattice.
The effective {\it vibrational temperature} is defined as: 
\begin{align}\label{eq:T}
T_{\bq\nu}(t) = \hbar \omega_{\bq\nu} \{k_{\rm B} \ln [ 1 + n_{\bq\nu}(t) ]\}^{-1},  
\end{align}
and it is obtained by inverting the Bose-Einstein distribution, Eq.~\eqref{eq:BE}.
At variance with $n_{\bq\nu}$, $T_{\bq\nu}$ becomes constant throughout the BZ 
at thermal equilibrium, and it is therefore better suited 
(but otherwise equivalent) to inspect the nonequilibrium dynamics of the lattice. 
Interpretation of $T_{\bq\nu}$ as a {\it thermodynamic temperature},  however, 
is rigorously justified only at thermal equilibrium. 
Figures~\ref{fig:fig_n}~(a-e)  report the average vibrational temperature 
$\tilde  T_{\bq} =  N_{\rm ph}^{-1} \sum_\nu T_{\bq\nu}$ -- with 
$N_{\rm ph}=9$ being the number of phonon modes of monolayer MoS$_2$ --  
for crystal momenta within the first BZ and 
for selected time steps. 
The same color bar (shown beside panel (i)) is used for panels (a-i). 

At $t=0$ (Fig.~\ref{fig:fig_n}~(a)), the lattice is at thermal equilibrium, 
as reflected by the constant vibrational temperature in the BZ 
($\tilde T_{\bq} = T^0_{\rm ph} = 100$~K). 
As the coupled electron-phonon dynamics begins ($t>0$), 
the excited carriers in the valence and conduction bands 
relax back to Fermi level by transferring energy to 
the lattice through the emission of phonons. 
The influence of these processes on the phonon distribution function  
is accounted for by the phonon-electron collision integral ($\Gamma^{\rm pe}_{{\bf q}\nu}$) 
in Eq.~\eqref{eq:bte_n}, which leads to an increase of $n_{{\bf q}\nu}$ (and thus of  $T_{\bq\nu}$)
as phonons with matching crystal momenta ${\bf q}$ and index $\nu$ are emitted.
After $t=100$~fs (Fig.~\ref{fig:fig_n}~(b)) the lattice has abandoned
the initial thermalized state, as revealed by the emergence of 
hot-spots in the BZ characterized by a higher average vibrational temperature $\tilde T_{\bq}$. 
In particular, an increase of the vibrational temperature is observed
for momenta close to $\Gamma$ and K, which in turn reflects an 
enhancement of the phonon population. 

To understand the origin of these features, 
it should be noted that the emission of phonons -- and, thus, the change of $T_{\bq\nu}$ --  
is triggered by electronic transitions within the valence and conduction bands, 
which are heavily constrained by energy and momentum conservation laws. 
For the excited electronic distribution of Fig.~\ref{fig:fig_f}~(a), for instance, 
phonon-assisted transitions within the valence band 
primarily involve two types of processes: 
(i) {\it intra-valley} transitions, connecting initial and final electronic 
states both located close to the same high-symmetry point ($\Gamma$ or K); 
(ii) {\it inter-valley} transitions, with the initial and final electronic states 
located at $\Gamma$ and K, respectively (or {\it vice versa}). 
Phonon-assisted transitions across the gap are forbidden by energy conservation. 
Due to momentum conservation, processes of type (i) result in the emission 
of long-wavelength phonons ($\bq \simeq 0$) with momenta 
close to $\Gamma$, whereas processes of type (ii) 
can only involve the emission phonons with momenta around K. 
A similar picture applies to transitions in the conduction band. 
Here, however, the presence of the Q valley also enables the emission 
of phonons around M and Q. 
A schematic illustration of the allowed inter- and intra-valley 
phonon-assisted transitions is provided in Fig.~S2 of the Supporting Information. 
Umklapp processes are also included in this pictures, 
since transitions connecting different BZs 
can be folded back to the first BZ via translation by 
a reciprocal lattice vector. 
This picture enables us to attribute the anisotropic 
increase of vibrational temperature to the preferential 
emission of phonons at $\Gamma$ and K, which is dictated 
by momentum selectivity in the electronic transitions.  

This mechanism leads to a further enhancement of the 
temperature anisotropy in the BZ 
for $t=500$~fs (Fig.~\ref{fig:fig_n}~(c)).
Additionally, an increase in vibrational  
temperature is observed at the M point and, less pronouncedly, at Q, 
which arise from transitions involving the 
Q pocket in the conduction band. 
On longer time scales, phonon-phonon scattering 
-- accounted for by the phonon-phonon collision 
integral ($\Gamma^{\rm pp}_{{\bf q}\nu}$) in Eq.~\eqref{eq:bte_n} -- 
counterbalances a non-thermal vibrational state 
by driving the lattice towards a thermalized 
regime (namely, $T_{\bq\nu} = $~constant). 
This behaviour is manifested for $t= 1.5$ and 3~ps
(Figs.~\ref{fig:fig_n}~(d-e)) by a progressive reduction 
of the temperature anisotropy in the BZ. 

The mode- and momentum-resolved 
vibrational temperature $T_{\bq\nu}$ -- superimposed to 
the phonon dispersion  in Figs.~\ref{fig:fig_n}~(g-i) for $t=0.1$, 0.5, and 3 ps --
may further change significantly for different phonon 
branches, since the contribution of each
phonon to the relaxation process is dictated by its  own
electron-phonon coupling strength.\cite{Darancet,Stern2018}
In particular, the stronger coupling of optical phonons 
makes them a more likely decay channel for the relaxation of excited carriers, as compared to 
other vibrations. 
This trend is reflected in Figs.~\ref{fig:fig_n}~(g-h) by the higher 
vibrational temperature of these modes throughout the initial 
stages of the dynamics, suggesting that the electronic 
coupling to optical modes plays a primary role in the 
emergence of non-thermal state of the lattice. 

A comprehensive picture of the formation and 
decay of a non-thermal vibrational state 
is provided in Fig.~\ref{fig:fig_n}~(f), which illustrates 
the time evolution of the average vibrational temperature 
$\tilde T_{\bq}$ for momenta along the M-K-$\Gamma$-K-M 
path in the BZ (dashed line in Fig.~\ref{fig:fig_n}~(a)). 
The most striking manifestation of the 
nonequilibrium lattice dynamics are clearly 
visible up to $t=5$~ps, even though weaker 
anisotropies in the vibrational temperature persist 
for 10~ps  or longer. 
The final temperature ($T_{\rm ph}^{\rm fin} \simeq 180$~K) reached after 
$t\simeq 10$~ps coincides with the final electronic temperature $T_{\rm el}^{\rm fin}$, 
marking the recovery of thermal equilibrium between electrons and phonons.  
The time dependence of the temperature anisotropy of the lattice
can be monitored by introducing the standard deviation $\sigma_T$:
\begin{align}\label{eq:std}
\sigma_T = \Omega_{\rm BZ}^{-1}[ 
   \langle \tilde  T_{\bf q}^2  \rangle_{\rm BZ}
-  \langle  \tilde T_{\bf q}  \rangle_{\rm BZ} ^2]^{\frac{1}{2}}
\end{align}
where $\langle \cdots \rangle_{\rm BZ}$
denotes the average over the Brillouin zone. 
In Fig.~\ref{fig:fig_n}~(j), values of $\sigma_T$ are reported for 
a concentration $n =1\cdot 10^{14}$~cm$^{-2}$ of excited carriers 
and several initial excitation energies (in units of meV per unit cell). 
The dot-dashed line corresponds to the same initial condition of panels (a-i).
$\sigma_T$ vanishes at thermal equilibrium ($t=0$), 
and it increasingly differs from zero for larger deviation from a thermalized state. 
This trend leads to a well-defined maximum in $\sigma_T$ at 
time delays between 200 and 900~fs, which marks 
the maximum anisotropy in the vibrational temperature. 
As more energy is provided to the initial electronic state, 
more time is required for the electrons to transfer their 
excess energy to the lattice via electron-phonon 
interactions, thus retarding the build-up of a 
nonequilibrium vibrational state. 
For all initial excited states, the return to a thermalized state takes place 
over timescales of the order of 5-10~ps. 

Overall, this picture enables us to clearly 
identify different stages in the out-of-equilibrium 
dynamics of electrons and phonons in monolayer MoS$_2$:
(i) following photo-excitation, electrons thermalize with the lattice
within 2~ps via electron-phonon scattering; 
(ii) on time-scale shorter than 1~ps, 
momentum selectivity in the phonon-assisted electronic transitions 
induces a highly-anisotropic population of different phonons modes, 
driving the lattice into a non-thermal vibrational state; 
(iii)  such non-thermal state persists for 5-10 picoseconds; 
(iv) thermal equilibrium is subsequently re-established by phonon-phonon scattering. 

Remarkably, owing to the inherent mismatch between the characteristic 
time-scales for the electronic relaxation via phonon 
emission ($<2$~ps) and lattice thermalization ($>5$~ps), 
the mechanism that underpins the emergence of 
a non-thermal vibrational states is very robust 
upon changes of the initial excitation conditions (Fig.~\ref{fig:fig_n}~(j)). 
This finding suggests that the 
transient enhancement of the phonon population at
few selected high-symmetry points in the Brillouin zone
could constitute a general characteristic of the 
nonequilibrium lattice dynamics of multi-valley semiconductors. 
Owing to the sensitivity of the electron-phonon interaction
on the phonon population 
one can postulate that, under suitable
condition of photo-excitation, the enhanced phonon population triggered
by the nonequilibrium dynamics of the lattice may ultimately
underpin transient changes of the electron-phonon interaction.
If this scenario could be realized, it could reveal new directions
towards the active control of many-body phenomena in quantum matter.

In conclusion, 
the coupled nonequilibrium dynamics of electrons and phonons in monolayer 
MoS$_2$ has been investigated 
by explicitly accounting for the effects of electron-phonon and phonon-phonon 
scattering within a first-principles time-dependent Boltzmann formalism. 
Owing to phase-space constraints in inter- and 
intra-valley electronic transitions, the decay path of excited
electrons and holes is dominated by the emission of
phonons with momenta around few selected high-symmetry points
in the Brillouin zone. 
Momentum selectivity in the phonon emission 
is found responsible for the emergence of a 
non-thermal vibrational state of the lattice which is 
characterized by a highly-anisotropic population of different 
phonon modes. 
Achieving control of non-thermal vibrational states and 
their lifetimes may provide unexplored opportunities to transiently 
tailor the electron-phonon coupling over time scales of several 
picoseconds. 

\begin{acknowledgement}
This project has been funded by the 
Deutsche Forschungsgemeinschaft (DFG) -- Projektnummer 443988403.  
\end{acknowledgement}

{\bf Supporting Information Available:} The Supporting Information is available free of charge. 

Computational details;\cite{hohenbergkohn,kohnsham1965,Giannozzi_2017,PhysRevB.45.13244,Pizzi_2020,bib:epw,ShengBTE_2014}
first-principles expressions for the collision integrals; 
details on the numerical implementation of Eqs.~\eqref{eq:bte_f} and \eqref{eq:bte_n}; 
starting-point dependent of the nonequilibrium lattice dynamics; 
schematic illustration of the allowed inter- and intra-valley transitions for electron-phonon scattering. 



\providecommand{\latin}[1]{#1}
\makeatletter
\providecommand{\doi}
  {\begingroup\let\do\@makeother\dospecials
  \catcode`\{=1 \catcode`\}=2 \doi@aux}
\providecommand{\doi@aux}[1]{\endgroup\texttt{#1}}
\makeatother
\providecommand*\mcitethebibliography{\thebibliography}
\csname @ifundefined\endcsname{endmcitethebibliography}
  {\let\endmcitethebibliography\endthebibliography}{}

\end{document}